\documentclass[oldversion]{aa}
\usepackage{txfonts}
\usepackage{graphicx}
\usepackage{natbib}
\bibpunct{(}{)}{;}{a}{}{,}

\def\xmm{{\it XMM-Newton}}
\def\intgr{{\it INTEGRAL}}
\def\bps{{\it BeppoSAX}}
\def\cra{{\it Chandra}}

\begin{document}
\title{\emph{INTEGRAL} Observations of the Perseus cluster}
\author{D. Eckert\inst{1} \& S. Paltani\inst{1}}
\offprints{Dominique Eckert, \email{Dominique.Eckert@unige.ch}}

\institute{$^1$ISDC Data Centre for Astrophysics, Geneva Observatory, University of Geneva, 16, ch. d'Ecogia, CH-1290 Versoix, Switzerland}
\date{Recieved / Accepted}

\abstract{We present the results of a 500 ksec observation of the Perseus cluster with \intgr, with the aim of investigating the possible diffuse non-thermal component detected in a previous \emph{Chandra} observation. In the 3-20 keV band with the JEM-X instrument, we detect the source with high significance and resolve it spatially. Above 20 keV with IBIS/ISGRI, we find that the source is point-like, and the cluster could be detected up to 120 keV. From the broad-band ISGRI/JEM-X spectrum, although we detect a non-thermal component, we find that the high-energy flux is variable and is consistent with the extrapolation of the 2-10 keV flux of the central AGN, NGC 1275. The extrapolation of the non-thermal component claimed from \emph{Chandra} data exceeds the \intgr\ spectrum by a factor of 3.}

\keywords{Galaxies: clusters: individual: Perseus Cluster - X-rays: galaxies: clusters - Gamma rays: observations}
\authorrunning{Eckert D. et al.}
\titlerunning{\emph{INTEGRAL} observations of the Perseus cluster}

\maketitle

\section{Introduction}

The Perseus cluster Abell 426 ($z=0.0176$) is the brightest galaxy cluster in the X-ray band, and the prototype of cooling-core clusters \citep{fabian}. Its central cD galaxy, NGC 1275, hosts a well-known narrow-line radio galaxy, Per A, which interacts with the intra-cluster gas through its jets and outflows \citep{boehringer}. In the X-ray band, the cluster was the target of a very long (900 ksec) \cra\ observation \citep{fabch1}, which revealed a very complex structure, with a temperature ranging from 2.5 keV in the central regions up to $\sim8$ keV in the outskirts. Several structures associated with the propagation of high-energy particles injected by the central AGN in the thermal plasma (X-ray cavities, sound waves) were also detected.\\

Apart from the very deep \emph{Chandra} observations, the cluster was also observed by \xmm\ \citep{churazov}. Thanks to the larger FOV of \xmm, it was also possible to observe the structure of the gas in the outer regions. An asymmetry of the surface brightness profile in the east-west direction was found, possibly corresponding to a small group of galaxies falling onto the main cluster. The contribution of the active nucleus in the center of NGC 1275 was also estimated. The spectrum of the AGN could be well-fitted by an absorbed power law with a photon index $\Gamma=1.65$ typical of radio-loud AGN and a luminosity of $10^{43}$ ergs $s^{-1}$ in the 0.5-8 keV band.\\

In the radio domain, the bright radio source 3C 84 is consistent with the position of NGC 1275. While most of the radio emission comes from the radio galaxy, emission on a larger scale ($\sim$10 arcmin) was also detected in the cluster (\citet{burns}, \citet{ferettiper}), designated as the Perseus ``mini-halo". This implies the presence of relativistic electrons, whose origin is not yet fully understood. If the AGN jets are hadronic, the radio emission probably comes from secondary electrons produced by interactions between cosmic-ray protons and thermal ions \citep{pfrommermh}. Alternatively, the electrons could also be directly accelerated through turbulence induced by the central nucleus \citep{gitti}, although the shocks produced by the interaction of the AGN jets with the ICM do not seem to be strong enough.\\

Since the existence of relativistic electrons in the cluster is obvious because of the radio synchrotron emission, we expect that photons of the CMB as well as optical/IR photons coming from the central galaxy should be up-scattered to higher energies, in particular to the (hard) X-ray domain (see e.g. \citet{sarazin}). An excess of emission compared to the thermal emission was probably detected in the Coma \citep{fusco} and A2256 \citep{fusco2256} clusters by \bps\ and in the Ophiuchus cluster \citep{eckertoph} by \intgr. Thanks to the very long \emph{Chandra} observation of the cluster, such a component has been claimed to be detected in addition to the thermal emission (\citet{sanders05}, \citet{sanders07}). This allowed the authors to present a magnetic field map of the cluster and derive a steep magnetic field profile, ranging from $\sim3\,\mu G$ in the center down to $\sim0.1\,\mu G$ in the outskirts. However, a recent \xmm\ observation did not confirm the result \citep{molendi}, so observations of the cluster above 10 keV where the thermal emission becomes weaker are required to confirm this result. \\

In the hard X-ray band, \citet{nevalainen} presented \bps/PDS observations of the cluster, and concluded that the emission was consistent with the extrapolation of the AGN flux. However, since the PDS instrument was non-imaging, it was not possible to put any constraints on the diffuse emission. Recently, \citet{ajello} found the same result using data from the \emph{Swift} satellite.\\

\begin{figure*}
\centerline{\hbox{\includegraphics[width=8cm]{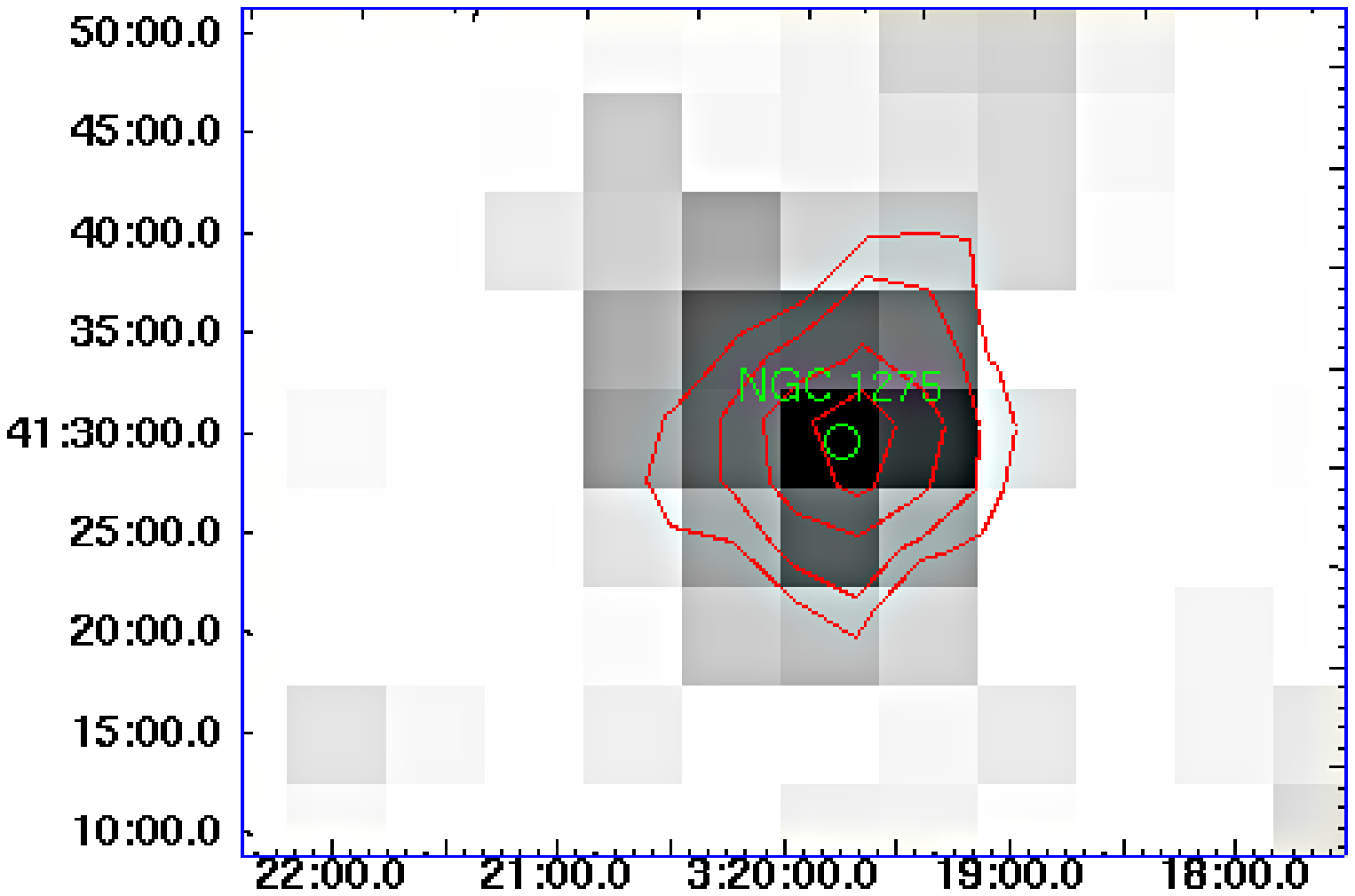}
\includegraphics[width=8cm]{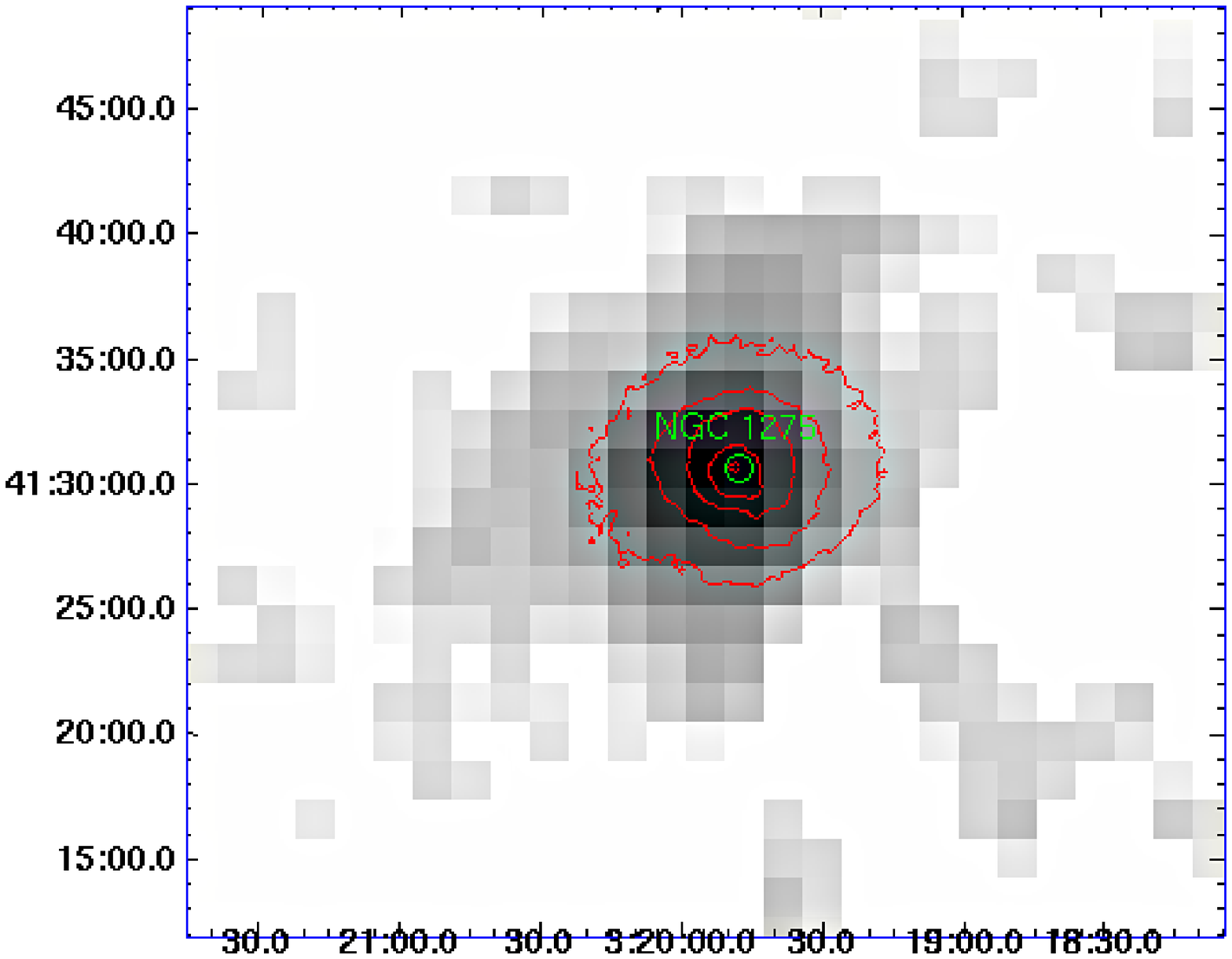}}}
\caption{Left: ISGRI 20-30 keV significance image of the Perseus cluster region. For comparison, the contours show the JEM-X 3-7 keV emission. The position of the radio galaxy NGC 1275 is also displaid. Right: JEM-X significance image of the cluster in the 3-7 keV band with contours from \emph{ROSAT}/PSPC. The green circle shows the position of NGC 1275 at the center of the cluster.}
\label{per_int}
\end{figure*}

In this paper, we present the results of a 500 ksec observation of the Perseus cluster with the \intgr\ satellite \citep{Win03}. Thanks to the broad-band coverage and good sensitivity in the hard X-ray range, IBIS \citep{Ube03} and JEM-X \citep{lund} have the necessary capabilities to constrain a possible non-thermal component. Indeed, the extrapolation of the diffuse non-thermal flux claimed by \citet{sanders05} within a radius of 3 arcmin from the center ($6.3\times10^{-11}$ ergs cm$^{-2}$ $s^{-1}$ in the 2-10 keV band, and a power-law index in the range $1.4-2.2$) is clearly above the sensitivity of IBIS. Therefore, if it is present, \intgr\ has the capabilities to confirm or rule out the \emph{Chandra} result.

\section{Data analysis}
\label{secdata}
\subsection{ISGRI data analysis}
\label{secisg}

The IBIS/ISGRI instrument on board \emph{INTEGRAL} \citep{lebrun} is a wide-field ($29^\circ\times29^\circ$) coded-mask instrument sensitive in the 15-400 keV band. Its angular resolution (12 arcmin FWHM nominal) is of the same order of magnitude as the size of the core of the cluster (core radius $R_c\sim 6'$, \citet{boehringer}). The Perseus cluster was observed during \intgr\ revolutions 0052, 0096, 0168 and 0220 for a total observing time of 500 ksec. During revolutions 0052 and 0096, the cluster was observed in staring mode. Unfortunately, this observing mode is not ideal for ISGRI, since artifacts due to an incorrect mask model become more prominent, and therefore the images deconvolved in the standard way are noisy. Nonetheless, the data in staring mode are fully usable for ISGRI spectral extraction, since the spectral-extraction tool works in the detector space and reduces considerably the parameter space for the deconvolution.\\

For ISGRI image reconstruction, we extracted detector shadowgrams using the standard OSA software v7.0 \citep{cour}. We fitted models of bright sources in the FOV to the raw data and subtracted them from the detector. After subtraction, we ignored improperly modeled pixels, and deconvolved the corrected images in the standard way (see \citet{eckertoph} for the details of the analysis method). Spectral extraction was performed using the standard OSA 7.0 software with a list of sources taken from the mosaic image. Given that the source is not spatially resolved, the instrument extracts the flux in a region of radius $\sim$6 arcmin corresponding to the half-width of the PSF.

\subsection{JEM-X data analysis}

The JEM-X X-ray monitor on board \intgr\ \citep{lund} consists of two identical X-ray detectors with coded mask, JEM-X1 and JEM-X2, sensitive in the 3-35 keV band. It is designed for the spectroscopic and imaging study of the sources detected by IBIS, with a better spatial resolution (3.35 arcmin FWHM nominal) and a field-of-view of $7.5^\circ$ half-response in diameter. Given that a large fraction of the observation was performed in staring mode, the JEM-X exposure time on Perseus is quite large (280 ksec).\\

JEM-X data analysis was performed in a completely standard way using the OSA 7.0 software. Images from JEM-X1 and JEM-X2 were combined to improve the signal-to-noise ratio. For JEM-X spectral extraction, we created mosaic images in 18 energy bands between 3 and 20 keV, we combined JEM-X1 and JEM-X2 mosaics and then extracted the JEM-X spectrum from the mosaic in a region of radius 6 arcmin similar to the IBIS PSF in order to ensure that both instruments view the same field.

\section{Morphology of the cluster}
\label{secmorph}

The source was clearly detected by both instruments at the $16\sigma$ (ISGRI, 20-30 keV) and 57$\sigma$ (JEM-X, 3-7 keV) level. Fig. \ref{per_int} shows the resulting significance images (i.e. ratio between the signal and the noise introduced by the background in each sky pixel) in the 20-30 keV band (ISGRI) and 3-7 keV band (JEM-X). The core radius of the cluster ($R_c\sim6$ arcmin) is comparable to the size of the IBIS PSF. However, fitting the ISGRI image, we find a half-width of $6.4$ arcmin, which is consistent with a point-like source. This could be due to the presence of the active nucleus in the center of the cluster, which might be responsible for a significant fraction of the emission above 20 keV.\\

On the other hand, the extension of the source is clearly seen in the JEM-X image. The total size of the source as seen by JEM-X is $\sim15$ arcmin, which is much larger than the PSF of the instrument (3.8 arcmin FWHM). As found by \xmm\ \citep{churazov}, the source is elongated in the East-West direction. In the 7-18 keV band, the source is also clearly detected, at the $21\sigma$ level. No significant change in the morphology is observed between the 2 bands.

\section{Broad-band \emph{INTEGRAL} spectrum}

In order to study the high-energy emission from the cluster and the central nucleus, we extracted the spectrum of the source in the 17-120 keV band with ISGRI and 3-20 keV band with JEM-X using the methods described in Sect. \ref{secdata}.

\subsection{3-20 keV JEM-X spectrum}

\begin{figure}
\resizebox{\hsize}{!}{\includegraphics[angle=270]{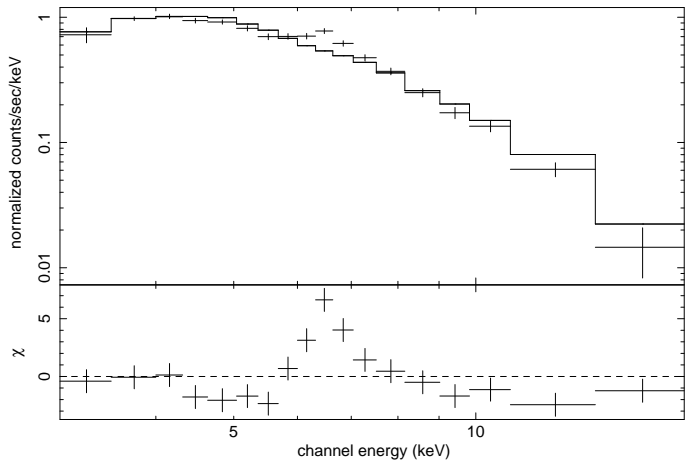}}
\caption{JEM-X spectrum of Perseus in the 3-20 keV band fitted by a MEKAL model with abundance fixed to 0. The residuals from the model are displayed in the bottom panel. A prominent emission line is detected around 6.5 keV.}
\label{per_jmx_fullres}
\end{figure}

Figure \ref{per_jmx_fullres} shows the JEM-X spectrum of Perseus in the 3-20 keV band. The spectrum was fitted by a MEKAL model at a temperature $kT=4.3$ keV. In order to emphasize the detection of an Fe emission line, the abundance was fixed to 0 for the fit. The spectral resolution of JEM-X is not sufficient to distinguish between the Fe XXV (at 6.5 keV) and Fe XXVI (6.8 keV) emission lines, so the detected line is a blend of the two. Introducing the abundace as a free parameter in the fit, we find a temperature $kT=3.84\pm0.13$ keV and an abundance of $0.37\pm0.05$ with respect to the solar value. The redshift of the cluster was fixed to $z=0.0176$ \citep{churazov}. The fit is satisfactory ($\chi^2=11.7$ for 15 d.o.f.).\\

From \xmm\ data, \citet{churazov} derived a temperature profile increasing from 3 keV in the central regions up to $\sim7$ keV in the outskirts. Our JEM-X spectrum is extracted within a radius $\sim6$ arcmin from the center, where \xmm\ derived a temperature increasing from 3 keV up to 6 keV. Given that the surface brightness profile is peaked towards the center, the JEM-X temperature is fully consistent with this result. \citet{churazov} also found an abundance of $\sim0.4$, in agreement with the JEM-X value. As a conclusion, we find that the 3-20 keV JEM-X spectrum gives results which are fully consistent with previous studies, e.g. \xmm/EPIC.

\subsection{Total spectrum}

In order to study the high-energy emission of the cluster, we fitted the JEM-X and ISGRI spectra simultaneously using the XSPEC package \citep{xspec}. The cross-calibration between the two instruments is known to be very close to 1 \citep{jourdain}. In any case, allowing for a cross-calibration factor between 0.8 and 1.2 does not change qualitatively any of the results presented here, so from now on the cross-calibration factor is fixed to unity. The results of the fitting procedure with different spectral models are summarized in Tab. \ref{tabspec}.\\

\begin{figure}
\resizebox{\hsize}{!}{\includegraphics[angle=270]{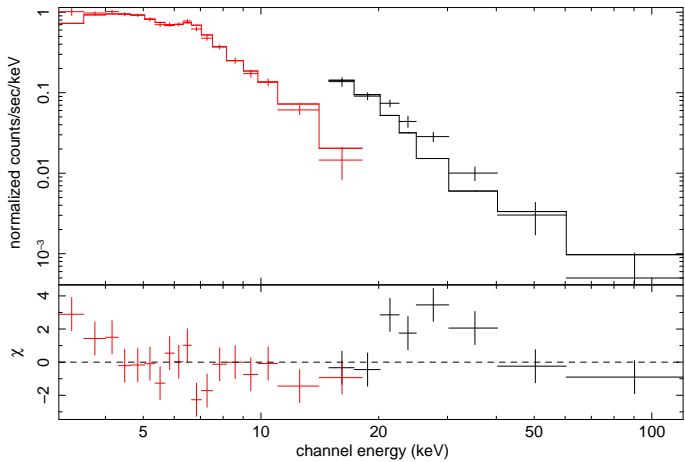}}
\caption{Total broad-band \intgr\ spectrum extracted with JEM-X (3-20 keV) and ISGRI (17-120 keV). The data were fitted using a single-temperature MEKAL model at $kT=4.1$ keV plus the contribution of the central nucleus of NGC 1275, which was fixed to the parameters derived by \citet{churazov}. While the high-energy emission is well-explained by this model, a significant excess ($4.8\sigma$) is found between 20 and 40 keV, indicating the need for another spectral component.}
\label{mekal+agn}
\end{figure}

As a first attempt to model the 3-120 keV spectrum, we fitted the spectrum with a single MEKAL model with abundance fixed to 0.4 solar and redshift fixed to $z=0.0176$ (model 1). The total spectrum is poorly represented by this model ($\chi^2_{red}=4.54$ for 24 d.o.f.), which clearly indicates the need for other spectral components. Adding a power-law component to the model (model 2), we find a significantly better fit ($\chi^2_{red}=1.15$ for 22 d.o.f.). The resulting power-law flux in the 2-10 keV band is $(1.6\pm0.2)\times10^{-10}$ ergs cm$^{-2}$ $s^{-1}$. This is more than an order of magnitude higher than the reported flux of NGC 1275 (e.g. \citet{donato}, \citet{churazov}). Moreover, hard X-ray emitting AGN normally show a much harder spectral index compared to the value $\Gamma=2.75_{-0.18}^{+0.16}$ derived by this model (see e.g. \citet{beckmann}), so this second component cannot be all associated with the central AGN. Fixing the contribution of the nucleus to the value found in \xmm\ data ($F_{2-10\mbox{\tiny{ keV}}}\sim10^{-11}$ ergs cm$^{-2}$ $s^{-1}$, $\Gamma=1.65$, (model 3), we find a significant excess ($4.8\sigma$) between 20 and 40 keV, while the high-energy data are consistent with the extrapolation of the AGN flux (see Fig. \ref{mekal+agn}). This means that a single thermal model plus the contribution from the AGN cannot explain the combined \intgr\ spectrum.\\

\begin{figure}
\resizebox{\hsize}{!}{\includegraphics[angle=270]{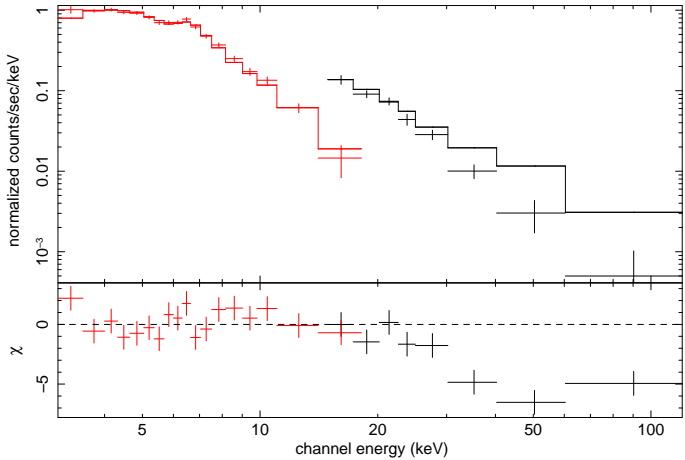}}
\caption{Total broad-band \intgr\ spectrum extracted with JEM-X (3-20 keV) and ISGRI (17-120 keV). The model (solid line) consists in a thermal component plus a power-law, with the parameters of the power-law as found by \citet{sanders05}. The model significantly over-predicts the data above 30 keV. Therefore, the presence of a power-law without cut-off at the level claimed by \citet{sanders05} is ruled out by \intgr\ data.}
\label{per_sanders}
\end{figure}

On the other hand, \citet{sanders05} reported on a possible diffuse non-thermal component with a 2-10 keV flux of $6.3\times10^{-11}$ ergs cm$^{-2}$ $s^{-1}$ and a photon index $\sim2.0$ (model 4). Adjusting a power-law model with such parameters in the \intgr\ spectrum clearly over-predicts ISGRI data over 30 keV, even when the AGN contribution is completely neglected. Figure \ref{per_sanders} shows the total \intgr\ spectrum with a model composed of a single thermal component plus a power-law component with the parameters from \citet{sanders05}. This model obviously over-predicts the high-energy data. The existence of a power-law component at the level claimed by \citet{sanders05} is therefore ruled out by ISGRI data, unless a high-energy cut-off is present in the spectrum.\\

\begin{table*}
\caption{Results of the fitting procedure to the joint ISGRI/JEM-X spectrum with different models.}
\label{tabspec}
\begin{center}
\begin{tabular}{c c c c c}
\hline
\hline
\# & Model & $\chi^2$/d.o.f. & $kT$ [keV] & Additional parameters\\
\hline
1 & MEKAL & 108.9/24 & 4.4 & \, \\
2 & MEKAL+power-law & 25.5/22 & $3.66\pm0.12$ & $\Gamma=2.75_{-0.18}^{+0.16}$ \\
3 & MEKAL+NGC 1275$^a$ & 55.9/24 & $4.1$ & \, \\
4 & MEKAL+NT$^b$ & 118.7/24 & $3.3$ & \, \\
5 & MEKAL+MEKAL+NGC 1275$^a$ & 23.9/22 & $3.41\pm0.15$ & $kT_{2}=11_{-3}^{+4}$ \\
6 & MEKAL+cutoffpl$^c$+NGC 1275$^a$ & 25.0/22 & $3.48\pm0.13$ & $E_{cut}=16_{-4}^{+6}$ \\
\hline
\end{tabular}
\end{center}
$^a$ Fixed to the value reported in \citet{churazov}\\
$^b$ Fixed to the parameters extracted by \citet{sanders05}\\ 
$^c$ Photon index of the cut-off power law fixed to $\Gamma=2.0$
\end{table*}

\begin{figure}
\resizebox{\hsize}{!}{\includegraphics[angle=270]{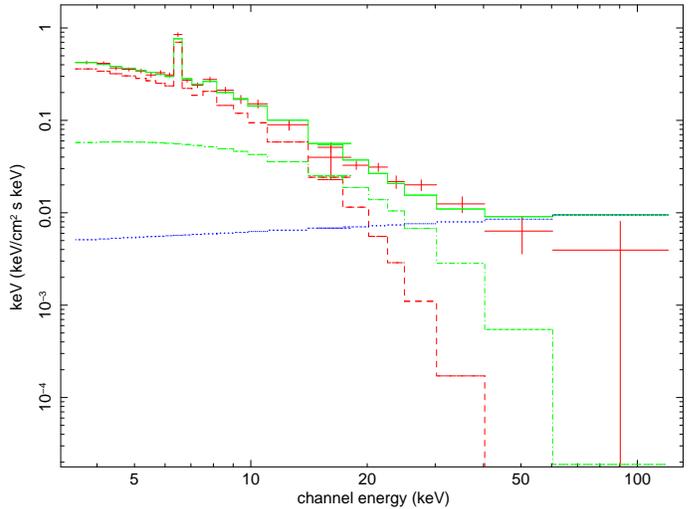}}
\caption{Total \intgr\ unfolded spectrum of the Perseus cluster. The green solid line shows the best fit to the data with a 3 keV MEKAL model (dashed red), a 7 keV bremsstrahlung model to describe the emission from the outer regions of the cluster (dashed green), plus the contribution of the central AGN NGC 1275 under the form of a power-law (dashed blue).}
\label{per_totalspec}
\end{figure}

Because of the complicated thermal properties of the gas in the inner 6 arcmin, modeling the data with a single-temperature thermal component cannot reproduce accurately the overall spectrum. In order to take into account the higher temperature found in the outer regions of the cluster ($kT\sim7$ keV from \xmm\ data), we added a second thermal component to the spectrum (model 5). Including the extrapolation of the AGN contribution, we get an excellent fit with a second thermal component at a temperature $kT=11_{-3}^{+4}$ keV. Figure \ref{per_totalspec} shows the total \intgr\ spectrum of Perseus in an $EF_{E}$ representation modeled with a two-temperature plasma plus the contribution of the AGN, fixed to the parameters derived by \citet{churazov}. The central temperature $kT=3.41\pm0.15$ keV is in agreement with \xmm\ data. The 2-10 keV relative contribution of the high-temperature component from the outer regions compared to the 3 keV component from the center is $\sim$15\%. From the JEM-X surface brightness profile, we find that the ratio between the integrated flux in the inner 4 arcmin and that in the 4-6 arcmin shell gives a similar value, so we expect that the cool component comes from the inner 4 arcmin, while the hotter component originates from the 4-6 arcmin shell. On the other hand, if instead of an additional thermal component we use a cut-off power-law with fixed photon index $\Gamma=2.0$ (model 6), we get an equally good fit with a cut-off energy $E_{cut}=16_{-4}^{+6}$ keV. In this case, the 2-10 keV flux of the cut-off power-law component is $9.1\times10^{-11}$ ergs cm$^{-2}$ $s^{-1}$. This is 1.5 times higher than the power-law flux reported by \citet{sanders05}, but given that it is integrated in a larger radius ($6'$ versus $3'$) the flux is consistent.\\

\section{Variability of the high-energy emission}
\label{varia}

As it is shown above, even with the addition of a second thermal component we find that a power-law is required to fit the high-energy data. However, this component is consistent with the extrapolation of the X-ray emission from the central nucleus. To confirm this, we investigated the variability of the flux at high energies. Indeed, any sign of variability would indicate a strong contamination of the high-energy flux by the AGN. Since the observations of the cluster were split into 4 different time periods, we extracted the 18-30 and 30-120 keV ISGRI fluxes for all 4 time periods (\intgr\ revolutions 0052, 0096, 0168 and 0220) and constructed light curves in these 2 bands. In the 30-120 keV band, we found obvious signs of source variability. Indeed, while the source was clearly detected in rev. 0168 at a flux level of $3.9\times10^{-11}$ ergs cm$^{-2}$ $s^{-1}$, no signal was detected during rev. 0096. Fitting the light curve with a constant flux value, we got a very bad fit, $\chi^2_{red}=10.4$ for 3 d.o.f., so the variability of the high-energy flux is significant. In the 18-30 keV band, a significant flux drop is seen during rev. 0096. Since the AGN contribution seemed to be negligible during this time period, this allowed us to measure the relative contribution of variable versus persistent flux in the lower band, which we found to be $\sim$35\%. Overall, our analysis reveals very clear signs of variability in ISGRI data, which allows us to conclude with good confidence that the power-law detected in the \intgr\ spectrum is due to the presence of the active nucleus in the center of NGC 1275.\\

\section{Discussion}

Analyzing a 500 ksec \intgr\ observation of the Perseus cluster, we have extracted a high signal-to-noise spectrum of the cluster in the 17-120 keV band with ISGRI and 3-20 keV band with JEM-X. Since the Perseus cluster hosts a variety of known thermal (cool core, complicated temperature structure) and non-thermal phenomena (AGN activity from the central galaxy, diffuse emission from the mini-halo), modeling the total spectrum of the cluster is a complicated task. \\

Using a very deep \emph{Chandra} observation of the cluster, \citet{sanders05} claimed the existence of diffuse non-thermal emission, which could be due to inverse-Compton scattering of the relativistic electrons responsible for the radio halo with CMB and infra-red photons. Extrapolating the non-thermal component in the hard X-ray band yields a flux which is clearly detectable by \intgr, so a confirmation of this result would be important. However, we find that the extrapolation of the power-law component clearly over-predicts the spectrum above 30 keV. Moreover, we found clear evidence that the high-energy flux ($E>30$ keV) is variable over a time-scale of several months, which proves that the origin of the high-energy emission cannot be diffuse. So based on \intgr\ data, we conclude that the existence of a power-law component without cut-off at the level claimed by \citet{sanders05} is ruled out.\\

On the other hand, assuming that the emission could be represented by a single-temperature model plus the contribution of the Seyfert nucleus does not give a good representation of the data. Therefore, an additional component in the form of a second thermal component or of a cut-off power-law is required. We find that both models can reproduce the data accurately, so based on \intgr\ data it is not possible to distinguish between these two scenarios. However, it is well-known from \emph{Chandra} and \xmm\ observations that the temperature structure of the cluster is complicated. In particular, \xmm\ data \citep{churazov} reveal a temperature of $\sim$3 keV in the inner 3 arcmin and a significantly higher temperature ($\sim 7$ keV) in the outskirts. Since ISGRI integrates the flux within a region of 6 arcmin from the cluster center, it is reasonable to fit the data with an additional thermal component at a higher temperature. If the additional component is a cut-off power-law, our data imply a high-energy cut-off at $E\sim16$ keV. This could be consistent with the statement of \citet{sanders07}, who discuss the possibility of a spectral break around 10 keV due to the cooling-time of non-thermal electrons at this energy.\\

In conclusion, although we cannot rule out the possibility that the emission is non-thermal and cuts off at $E\sim16$ keV, the fit with two temperatures of $\sim3$ keV and $\sim10$ keV is satisfactory. It also appears quite natural given that the source exhibits multiple thermal components in the inner 6 arcmin, in particular a 3 keV component from the cool core and a 7 keV component from the outer regions. Therefore, we conclude that the total \intgr\ spectrum can be perfectly well described by a thermal model consistent with previous studies, plus a variable power-law component from the central AGN. Recently, the result of \citet{sanders05} was not confirmed by a longer \xmm\ observation \citep{molendi}. The authors claim that the discrepancy is due to uncertainties in the effective area calibration of \emph{Chandra} in the highest band, which would strongly affect the results. Indeed, since the band-pass of \emph{Chandra} stops at 8 keV, it is difficult to distinguish between a high-temperature bremsstrahlung and a power-law emission. The flux difference between a $\Gamma=2.0$ power-law model and a $kT=7.0$ keV bremsstrahlung model with the same 2-10 keV flux in the 7-8 keV band is 2.5\%, hence the combination of the statistical and systematic uncertainties should be better than 2.5\% in the 7-8 keV band to distinguish between the 2 models.\\

Nevertheless, the presence of the radio mini-halo clearly indicates the existence of relativistic electrons not associated with the central AGN. Therefore, IC scattering of the high-energy electrons with CMB and IR photons should occur, and a diffuse non-thermal component should exist as well in the X-ray domain. However, in the center of strong cooling-core clusters like Perseus, the density of the gas is very high, and the shocks induced by the interaction between the AGN radio lobes and the thermal plasma amplify the magnetic field. Measurements of magnetic fields through Faraday rotation measure in cooling-core clusters (see e.g. \citet{carilli}) indicate high magnetic field values, $B\sim10\,\mu G$. If the magnetic field in the center of Perseus is of the order of $10\,\mu G$, the electrons should cool mostly through synchrotron processes, and the level of IC emission should be lower. Using high magnetic field estimates together with the radio flux of the Perseus mini-halo, \citet{pfrommer3} predicts a 2-10 keV non-thermal flux of $\sim5\times10^{-13}$ ergs cm$^{-2}$ $s^{-1}$, over 2 orders of magnitude lower than the flux claimed by \citet{sanders05}, and clearly below the sensitivity limit of \intgr/ISGRI. This would explain the non-detection of diffuse non-thermal emission by \intgr.

\section{Conclusion}

In this paper, we presented the results of a 500 ksec \intgr\ observation of the Perseus cluster. We found that \intgr\ data can efficiently constrain a possible high-energy tail in the spectrum. Although we found clear evidence for a power-law component above 30 keV, the properties of this power-law are consistent with the extrapolation of the X-ray spectrum from the central nucleus, in agreement with the results from \bps\ \citep{nevalainen} and \emph{Swift} \citep{ajello}. Moreover, we have found that the high-energy flux (30-120 keV) significantly varies between the different periods of the observation, covering $\sim$ 1.5 yr. This clearly demonstrates that the high-energy emission is dominated by the AGN.\\

On the other hand, extrapolating the power-law component claimed by \citet{sanders05} from \emph{Chandra} data ($6.3\times10^{-11}$ ergs cm$^{-2}$ $s^{-1}$ in the  2-10 keV band with a photon index $\Gamma\sim2.0$) to the ISGRI band, we found that this model clearly over-predicts the high-energy data by a factor of $\sim$3 in the 40-60 keV band. Therefore, the presence of a flat power-law at such a high level is ruled out by ISGRI data. Overall, we find that the broad-band \intgr\ spectrum can be well-described by a two-temperature thermal model (a $\sim$ 3 keV component from the cool core and a hotter component from the outskirts) plus a variable power-law from the central radio galaxy. The data can also be well-described by a model consisting of a single thermal component, the AGN contribution and a power-law with a high-energy cut-off at an energy $E_{cut}\sim16$ keV in agreement with the discussion of \citet{sanders07}. Given that deep X-ray observations clearly indicate the multi-temperature structure of the cluster, the two-temperature model is favoured. Moreover, a longer \xmm\ observation \citep{molendi} did not confirm the \emph{Chandra} result. The authors claim that the discrepancy between \emph{Chandra} and \xmm\ results is due to uncertainties in the effective area calibration of \emph{Chandra} in the highest band. \intgr\ data are in agreement with this result.\\

In conclusion, using \intgr\ data, we found no evidence for a diffuse power-law emission which would dominate the emission above 30 keV. Unfortunately, the angular resolution of IBIS/ISGRI is not sufficient to disentangle the point-like emission from the diffuse component, so it is not possible to set any upper limit on the diffuse non-thermal emission. Observations of the cluster in the hard X-ray band with better angular resolution and sensitivity would be very important to detect the diffuse non-thermal component and map the magnetic field strength over the cluster, which could bring a very important input to the understanding of the interactions between the AGN radio lobes and the thermal plasma. In this framework, the future focusing hard X-ray missions (\emph{Simbol-X}, \emph{NuSTAR}, \emph{NeXT}) should be able to detect the diffuse non-thermal component.

\begin{acknowledgements}
This work is based on observations with INTEGRAL, an ESA project with instruments and science data center funded by ESA member states (especially the PI countries: Denmark, France, Germany, Italy, Switzerland, Spain), Czech Republic and Poland, and with the participation of Russia and the USA.
\end{acknowledgements}

\bibliographystyle{aa}
\bibliography{aa081154}

\end{document}